\documentclass[twocolumn,amsmath]{revtex4-1}

\usepackage{graphicx}% Include figure files

\usepackage[colorlinks=true,linktocpage=true,linkcolor=blue,citecolor=blue]{hyperref}

\begin{document}

\title{Production of $X_{cs\bar{c}\bar{s}}$ in heavy ion collisions}

\author{Yuanyuan Hu}
\affiliation{Guangdong Provincial Key Laboratory of Nuclear Science, Institute of Quantum Matter, South China Normal University, Guangzhou 510006, China}
\affiliation{Guangdong-Hong Kong Joint Laboratory of Quantum Matter, Southern Nuclear Science Computing Center, South China Normal University, Guangzhou 510006, China}

\author{Hui Zhang}\email{Mr.zhanghui@m.scnu.edu.cn}
\affiliation{Guangdong Provincial Key Laboratory of Nuclear Science, Institute of Quantum Matter, South China Normal University, Guangzhou 510006, China}
\affiliation{Guangdong-Hong Kong Joint Laboratory of Quantum Matter, Southern Nuclear Science Computing Center, South China Normal University, Guangzhou 510006, China}

\date{\today}

%%%%%%%%%%%%%%%%
\begin{abstract}
The yields of $X_{cs\bar{c}\bar{s}}$ with its two possible configurations, i.e., the hadronic molecular state and tetraquark state, for Pb-Pb collisions at $\sqrt{s_{NN}}=5.02~\rm{TeV}$ is studied. A volume effect is found from the centrality distribution of $X_{cs\bar{c}\bar{s}}$, which could help to distinguish the inner structure of $X_{cs\bar{c}\bar{s}}$. We also show the rapidity and the transverse momentum distributions of $X_{cs\bar{c}\bar{s}}$ production as well as its elliptic flow coefficient as a function of the transverse momentum.
\end{abstract}

%\pacs{}
\keywords{Exotic}
\maketitle

\section{Introduction}

Quarks and gluons are the fundamental degrees of freedom of quantum chromodynamics (QCD). Because of the nonperturbative feature of QCD, we can only observe confined colorless hadrons. A normal hadron has two modes: a meson is made up of one quark and one antiquark, and a baryon is made up of three (anti)quarks. Multiquark hadrons made up of more than three quarks were proposed at the beginning of the construction of the quark model by Gell-Mann and Zweig \cite{Gell-Mann:1964ewy, Zweig:1964ruk, Zweig:1964jf}. However, the existence of tetraquarks and pentaquarks was not proven until the observation of $XYZ$ states \cite{ParticleDataGroup:2020ssz}, hidden-charm $P_c$ states \cite{LHCb:2015yax, LHCb:2019kea, LHCb:2020jpq, LHCb:2021chn}, doubly-charm $T_{cc}^+$ \cite{LHCb:2021vvq, LHCb:2021auc} and fully-charm tetraquark states \cite{LHCb:2020bwg}, etc. \cite{Chen:2016qju, Esposito:2016noz, Guo:2017jvc, Liu:2019zoy, Brambilla:2019esw, Chen:2022asf, Meng:2022ozq}

Five $J/\psi \phi$ structures $X(4140)$, $X(4274)$, $X(4500)$, $X(4685)$ and $X(4700)$ in the $B^+ \to J/\psi \phi K^+$ decay process were observed by the LHCb Collaboration \cite{LHCb:2016nsl, LHCb:2016axx, LHCb:2021uow}, CDF Collaboration \cite{CDF:2009jgo, CDF:2011pep}, CMS Collaboration \cite{CMS:2013jru}, D0 Collaboration \cite{D0:2013jvp} and BARAR Collaboration \cite{BaBar:2014wwp}. $X(4140)$ and $X(4274)$ are considered as the $cs\bar{c}\bar{s}$ tetraquark ground states, whereas $X(4500)$ and $X(4700)$ are considered as the $cs\bar{c}\bar{s}$ tetraquark excited states, in various theoretical methods \cite{Chen:2010ze, Stancu:2009ka, Chen:2016oma, Ebert:2008kb, Lu:2016cwr, Wu:2016gas, Liu:2021xje, Deng:2017xlb, Wang:2016gxp}. In Ref. \cite{Wang:2021ghk}, $X(4685)$ was also considered as the axial vector 2S radial excited $cs\bar{c}\bar{s}$ tetraquark state. In Refs. \cite{Reinders:1984sr, Shifman:1978bx, Turkan:2021ome, Yang:2022zxe}, the mass spectra of the S-wave and D-wave $cs\bar{c}\bar{s}$ tetraquarks in different excitation structures are calculated using the QCD sum rules method. There have been many theoretical studies on the inner structure of these $X$'s, such as the molecular states \cite{Liu:2009ei, Mahajan:2009pj, Wang:2009ue, Wang:2009ry, Albuquerque:2009ak, Branz:2009yt, Ding:2009vd, Zhang:2009st, Shen:2010ky, Liu:2010hf, Finazzo:2011he, He:2011ed, Hidalgo-Duque:2012rqv, Wang:2014gwa, Ma:2014ofa, Ma:2014zva, Karliner:2016ith, MartinezTorres:2016cqv, Mutuk:2022ckn}, compact or diquark-antidiquark states \cite{Stancu:2009ka, Drenska:2009cd, Ebert:2010zz, Chen:2010ze, Vijande:2014cfa, Patel:2014vua, Anisovich:2015caa, Wang:2015pea, Zhou:2015frp, Padmanath:2015era, Lebed:2016yvr}, cusp effects \cite{vanBeveren:2009dc, Luo:2022xjx, Nakamura:2021bvs}, dynamically generated resonances \cite{Molina:2009ct, Branz:2010rj}, conventional charmonium \cite{Liu:2009iw}, and hybrid charmonium states \cite{Mahajan:2009pj, Wang:2009ue}. However, overall, the inner structure of $X(4140)$, $X(4274)$, $X(4500)$, $X(4685)$ and $X(4700)$ remains an open question.

In the molecular picture, a $X_{cs\bar{c}\bar{s}}$ is formed by a strange-charmed meson $D_s^+$ ($D_s^-$) and a $D_s^{*-}$ ($D_s^{*+}$), while a $X(3872)$ is formed by a charmed meson $D_0$ ($D_0^*$, $D^+$, $D^-$) and a $\bar{D_0}$ ($\bar{D_0^*}$, $D^{*-}$, $D^{*+}$). In the tetraquark picture, a $X_{cs\bar{c}\bar{s}}$ is formed by a spin triplet diquark $[cs]_1$ (spin singlet diquark $[cs]_0$) and a spin singlet antidiquark $[\bar{c}\bar{s}]_0$ (spin triplet antidiquark $[\bar{c}\bar{s}]_1$), while a $X(3872)$ is formed by a diquark $[cq]_1$ ($[cq]_0$) and a $[\bar{c}\bar{q}]_0$ ($[\bar{c}\bar{q}]_1$), $q$ for $u$ and $d$ quarks. Although light quarks $u$ and $d$ in $X(3872)$ are replaced with $s$ quarks in $X_{cs\bar{c}\bar{s}}$, their inner structures may or may not be the same. This motivates the present study, in which we examine whether the approach we proposed in Ref.~\cite{Zhang:2020dwn} can also be applied to the $X_{cs\bar{c}\bar{s}}$ case and thus find a way to distinguish the two internal structures with heavy ion measurements.
In this work, we try to distinguish the two aforementioned possible inner structures of $X_{cs\bar{c}\bar{s}}$, i.e., a loose hadronic molecule or a compact tetraquark, by studying its production in heavy ion collisions. The remainder of this paper is organized as follows. In section~\ref{sec:Framework}, we introduce the generation mechanism of $X_{cs\bar{c}\bar{s}}$ into the AMPT model corresponding to its two possible inner structures following the production of $X(3872)$ described in Ref.~\cite{Zhang:2020dwn}. In section~\ref{sec:Results}, we examine the production of $X_{cs\bar{c}\bar{s}}$ as a function of centrality, transverse momentum, and rapidity. A volume effect is found, which can be a probe of the inner structure of $X_{cs\bar{c}\bar{s}}$. A summary and outlook are presented in section~\ref{sec:Summary}.

\section{Framework}\label{sec:Framework}

\begin{figure}[hbt!] 
   \includegraphics[height=4cm]{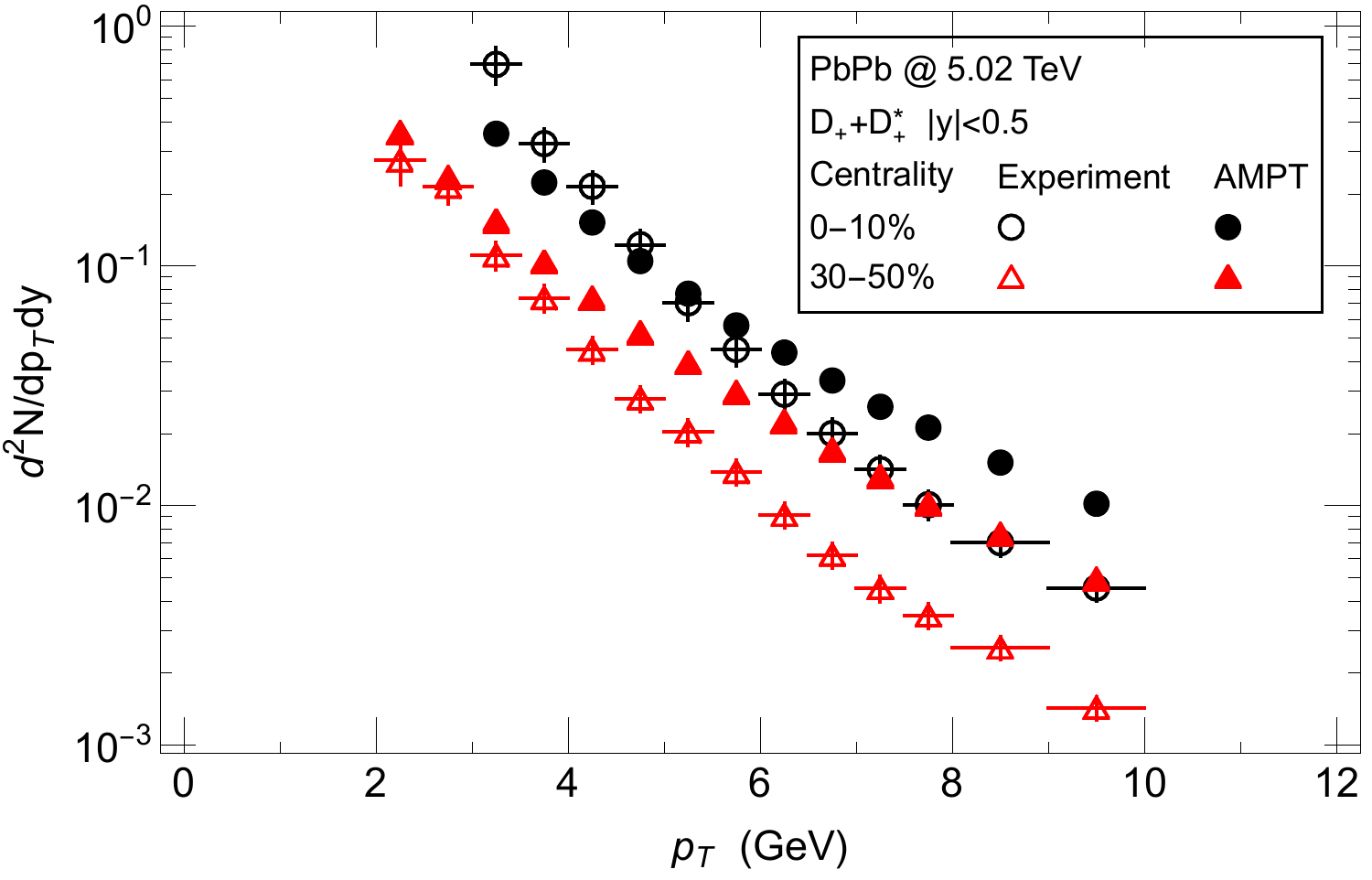}
   \includegraphics[height=4cm]{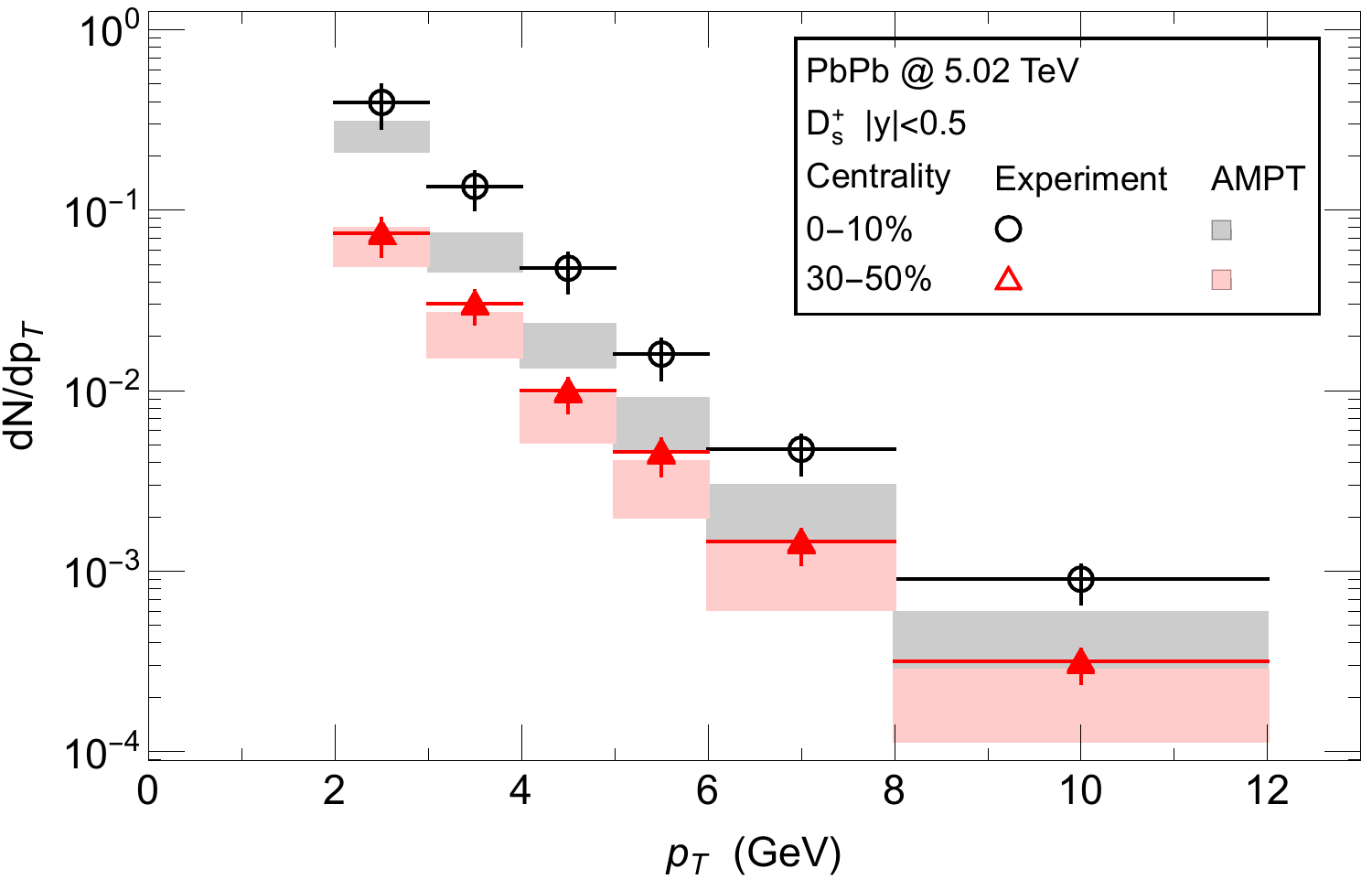}
   \caption{(color online) Upper panel: total production of $D^+ + D^{+*}$ from the ALICE Collaboration \cite{ALICE:2021rxa}; Lower panel: the production of $D_s^+$ from the ALICE Collaboration \cite{ALICE:2021kfc}. The bands reflect the uncertainty due to constituent composition as discussed around Eq.~\eqref{eq:thermal} that are obtained from varying the composition fraction by $\pm 10\%$.} 
   \label{fig:Exp_AMPT}
\end{figure}

In this study, we generate a total of one million minimum bias events for Pb-Pb collisions at $\sqrt{s_{NN}}=5.02~\rm{TeV}$ by using the framework developed in Ref.~\cite{Zhang:2020dwn}. We introduce the production mechanism to produce $X_{cs\bar{c}\bar{s}}$ for its two possible configurations, i.e., the hadronic molecular configurations and the tetraquark configurations into the default version (v1.26t9b) of AMPT transport model~\cite{Lin:2004en}. Given that $X_{cs\bar{c}\bar{s}}$ contains (anti-)charm quarks and (anti-)strange quarks, we need to generate a reasonable number of individual charm and strange quarks in the partonic phase. On top of the default version of AMPT, we modify the factor of $K$ (\cite{Zhang:2019utb}) to enhance the initial $c$ and $\bar{c}$ spectra because of a lack of some channels related to initial heavy quarks. The AMPT calculation gives a reasonable (order-of-magnitude) description of the experimental data \cite{ALICE:2021rxa} for the total yield of $D^+ + D^{+*}$ in the low $p_T$ region (see upper panel of Fig.~\ref{fig:Exp_AMPT}). For the strange quarks, an upper limit on the relative production of strange to non-strange quarks in AMPT is set to 0.6 because of the strangeness enhancement effect (see \cite{Lin:2014tya}), and our calculations also give a reasonable (order-of-magnitude) description of the experimental data \cite{ALICE:2021kfc} for the yield of $D_s^+$ meson (see lower panel of Fig.~\ref{fig:Exp_AMPT}). The main purpose of this work is to distinguish two inner structures of $X_{cs\bar{c}\bar{s}}$ through their significantly different production rates. The difference of $D$ and $D_s^+$ mesons production between our calculation and experimental data should not influence the relative yield between two inner structures and thus cannot change the qualitative results. 

We use the same production mechanism developed in Ref.~\cite{Zhang:2020dwn} for the hadronic molecule and tetraquark configurations of the $X_{cs\bar{c}\bar{s}}$. For the molecular picture, the charmed-strange mesons are collected after the hadronization process. Then, $D_s^+$ ($D_s^-$) and $D_s^{-*}$ ($D_s^{+*}$) are coalesced (similar to the hadronization process mentioned in~\cite{Lin:2004en}) to form the ``molecule" $X_{cs\bar{c}\bar{s}}$ according to the following conditions: the relative distance within the region $[5\rm{fm},~ 7\rm{fm}]$ and invariant mass within the region $[2M_{D_s^+},~ 2M_{D_s^{+*}}]$. For the tetraquark picture, the ``tetra" $X_{cs\bar{c}\bar{s}}$ is formed via two steps. (i) First, diquarks ($cs$) and diquarks ($\bar{c}\bar{s}$) are formed by matching a (anti-)charm quark with the nearest (in both position space and momentum space) (anti-)strange quark in the parton. (ii) Then, these (anti)diquarks are coalesced to form the $X_{cs\bar{c}\bar{s}}$ according to the following conditions: the relative distance $< 1\rm{fm}$ and invariant mass within the region $[2M_{[cs]_1},~ 2M_{[cs]_0}]$ (the spin triplet and singlet diquark masses are defined in Refs.~\cite{Ebert:2008kb, Lu:2016cwr}). Owing to a lack of spin information in the AMPT model for the formation of the charmed-strange mesons and (anti)diquarks, the relative yield ratios are estimated using the thermal model:
\begin{equation}
   R(\frac{A}{B})\equiv \frac{\text{Yield}(A)}{\text{Yield}(B)}= e^{-(m_A-m_B)/T_{\text{freezeout}}},
   \label{eq:thermal}
\end{equation}
where $m_A$ and $m_B$ represent the masses of hadrons A and B, respectively. Here, $T_{\text{freezeout}}\simeq160~\mathrm{MeV}$ is the freeze-out temperature. For the hadronic picture, A and B are the $D_s^+$ and $D_s^{+*}$ mesons, respectively. For the tetraquark picture, A and B are the spin triplet and singlet diquark, respectively. This estimate indicates a composition of $30\%$($70\%$) for $D_s^+$($D_s^{+*}$) and a composition of $35\%$($65\%$) for spin triplet(singlet) diquarks. We also vary the composition between $20\%$($80\%$) and $40\%$($60\%$) to show the uncertainty bands.

\section{Results and Discussions} \label{sec:Results}

Within this simulation framework, we use the Monte Carlo method to generate a total of one million minimum bias events for Pb-Pb collisions at $\sqrt{s_{NN}}=5.02\ \rm{TeV}$. The inclusive yield of $X_{cs\bar{c}\bar{s}}$ is found to be approximately 42000 in the molecular picture and approximately 200 in the tetraquark picture. As a benchmark for comparison, we also estimate the yield of $X(3872)$ within the same framework (see the production mechanism in Ref.~\cite{Zhang:2020dwn}, the yield should be multiplied a factor $\tfrac{1}{4}$ owing to wavefunction normalization for both the molecular and tetraquark pictures). The inclusive yield of $X(3872)$ is found to be approximately 171000 in the molecular picture and approximately 600 in the tetraquark picture. The yield of $X_{cs\bar{c}\bar{s}}$ is approximately $\tfrac{1}{4}$ of that of $X(3872)$. Compared with the experimental data of $X(3872)$ measured by the CMS collaboration for Pb-Pb collisions at $\sqrt{s_{NN}}=5.02\ \rm{TeV}$ \cite{CMS:2021znk}, our finding suggests that an observable signal of $X_{cs\bar{c}\bar{s}}$ could be measured in heavy ion collisions at the LHC energy.

One can also find the production in the molecular picture significantly exceeds that in the tetraquark picture, by a factor of 200 --- a 2-order-of-magnitude difference. This result may be understood as follows: the $c-\bar{c}$ and $s-\bar{s}$ quarks must be pair produced in the initial conditions of heavy ion collisions and then expand and cool with the bulk flow; the molecule $X_{cs\bar{c}\bar{s}}$ needs a large volume to be formed, while the tetraquark $X_{cs\bar{c}\bar{s}}$ needs a compact volume to be formed; thus, the probability of the formation of hadron molecules is far higher than that for the tetraquark state. 

\begin{figure}[hbt!] 
\includegraphics[height=5cm]{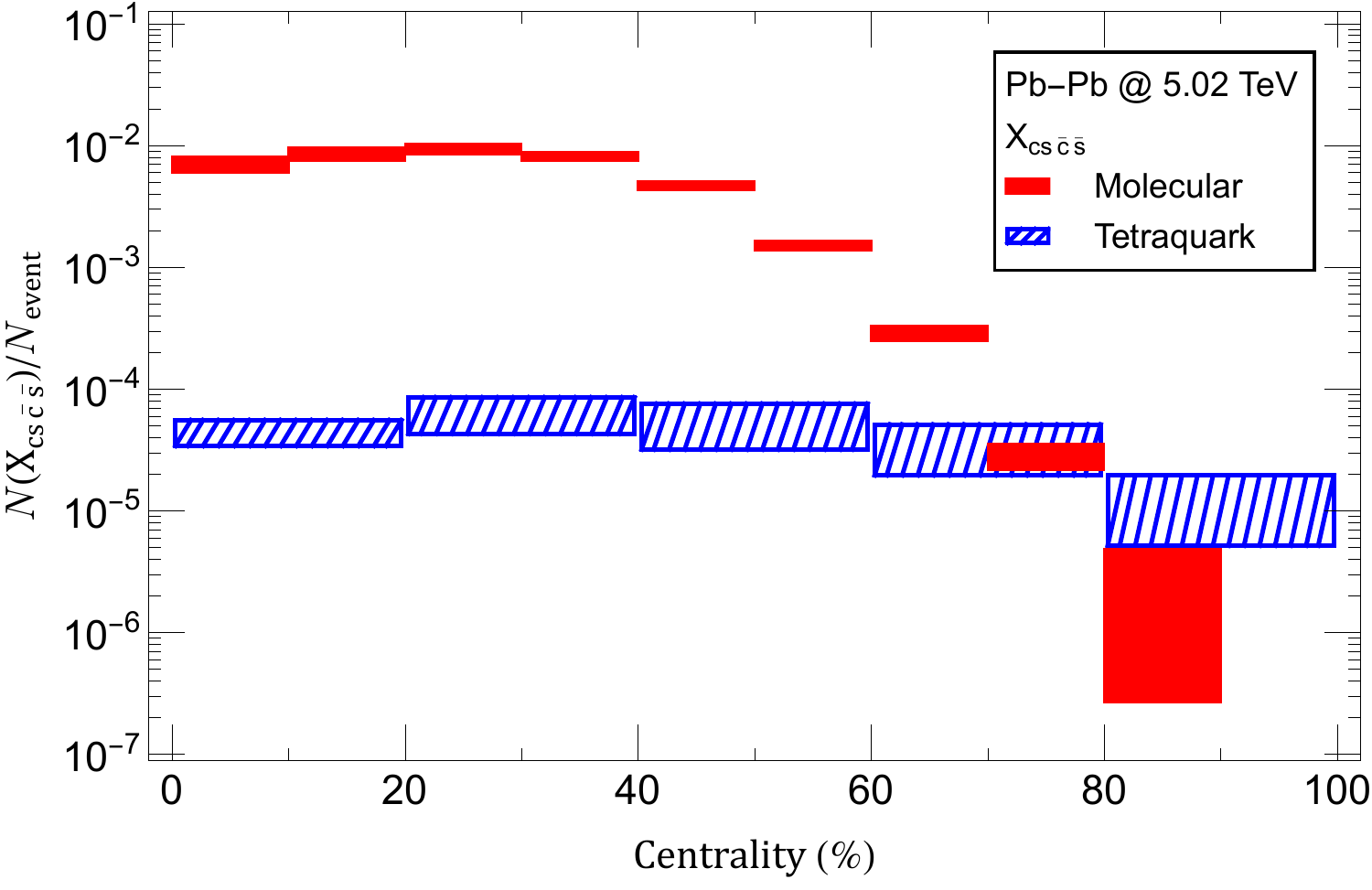}
\caption
{(color online) Centrality dependence of the $X_{cs\bar{c}\bar{s}}$ in Pb-Pb collisions at $\sqrt{s_{NN}}=5.02~ \rm{TeV}$ for hadronic molecular configuration (red solid boxes) and tetraquark configuration (blue shaded boxes). The bands reflect both statistical uncertainty from our simulations and the uncertainty due to constituent composition as discussed around Eq.~\eqref{eq:thermal}, and are obtained from varying the composition fraction by $\pm 10\%$.} 
\label{fig:centrality}
\end{figure}
We plot the $X_{cs\bar{c}\bar{s}}$ production as a function of centrality in Pb-Pb collisions at $\sqrt{s_{NN}}=5.02~ \rm{TeV}$ for the hadronic molecular state and tetraquark state in Fig.~\ref{fig:centrality}. One can find the yield of the $X_{cs\bar{c}\bar{s}}$ in the molecular picture is 2 orders of magnitude larger than that in the tetraquark picture. From the central collision region to the peripheral collision region, the production first increases then decreases for both the molecular state and the tetraquark state, and the slope of the decrease is far larger in the molecular state than in the tetraquark state. This results from a competing effect between the volume of the bulk system and the size of $X_{cs\bar{c}\bar{s}}$. For central collisions, the number of (anti-)charm and (anti-)strange quarks is large, the bulk volume is large, and its evolution time is long; thus, the (anti-)charm and (anti-)strange quarks separate sufficiently, which benefits the production of a large-size molecular state. For the peripheral collisions, both the number of (anti-)charm and (anti-)strange quarks and the size of the fireball are small; as such, the evolution time of the fireball is short, which benefits the production of small-sized tetraquark states. This size effect could help to explore the internal structure of $X_{cs\bar{c}\bar{s}}$ through different collision systems, e.g., Pb-Pb, Au-Au, Xe-Xe, Cu-Cu, O-O, and $d-A$/$p-A$.

\begin{figure}[hbt!] 
\includegraphics[height=5cm]{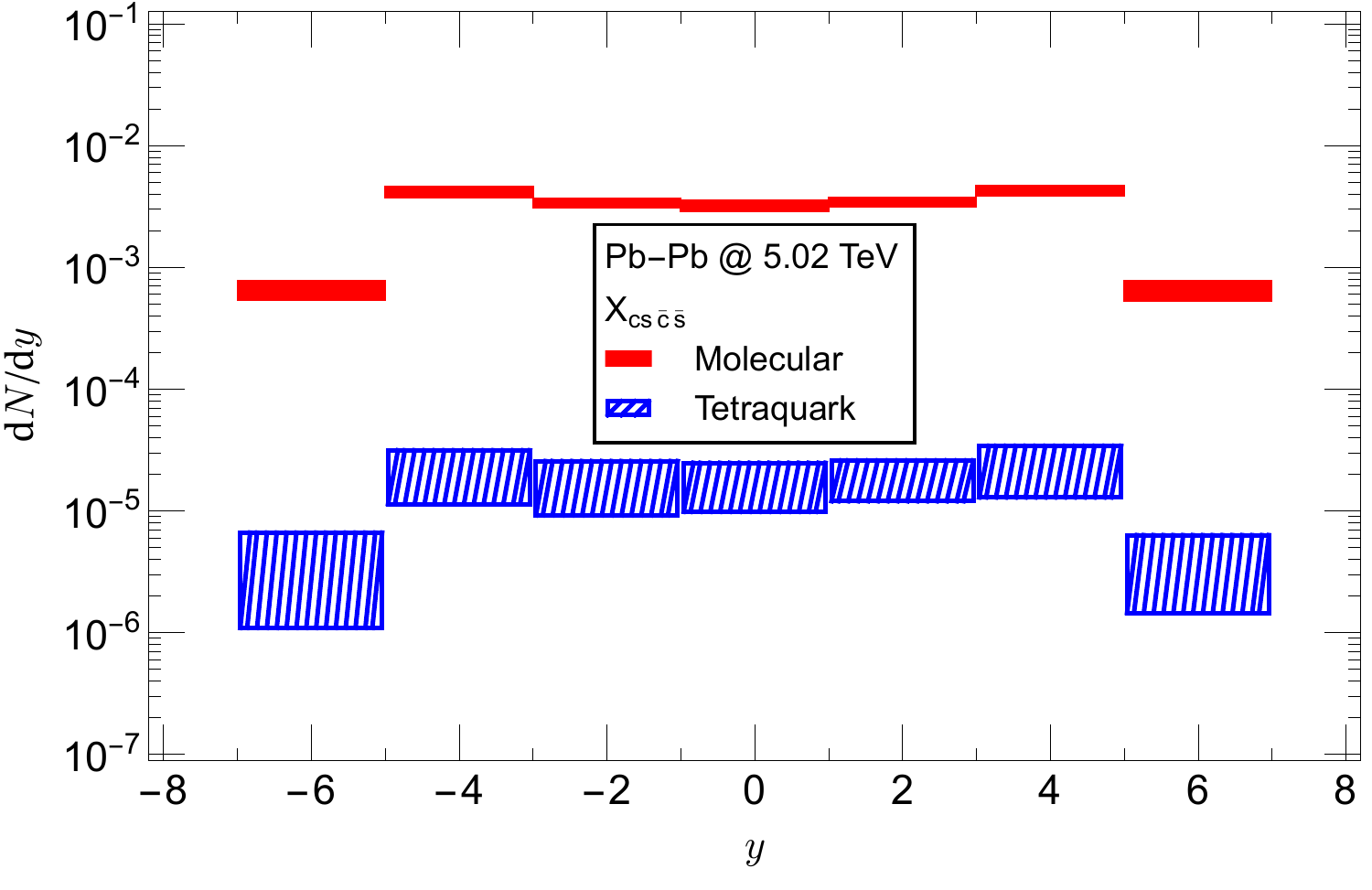}
\includegraphics[height=5cm]{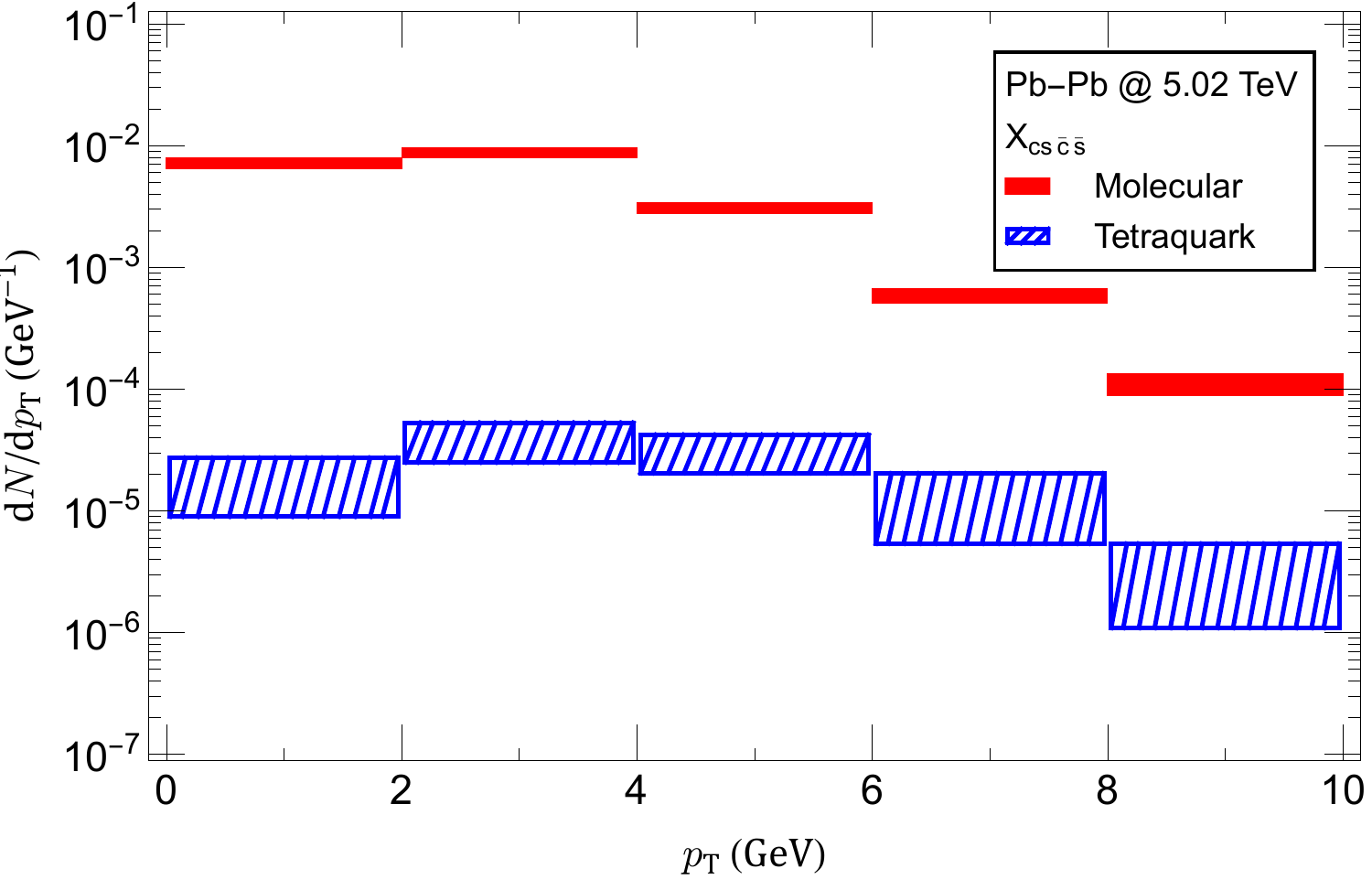}
\caption
{(color online) Rapidity $y$ and transverse momentum $p_T$ distribution of the $X_{cs\bar{c}\bar{s}}$ yield in Pb-Pb collisions at $\sqrt{s_{NN}}=5.02~\mathrm{TeV}$ for hadronic molecular configuration (red solid boxes) and tetraquark configuration (blue shaded boxes). The bands are determined as described in Fig.~\ref{fig:centrality}.  } 
\label{fig:rapidity}
\end{figure}
\begin{figure}[hbt!] 
\includegraphics[height=5cm]{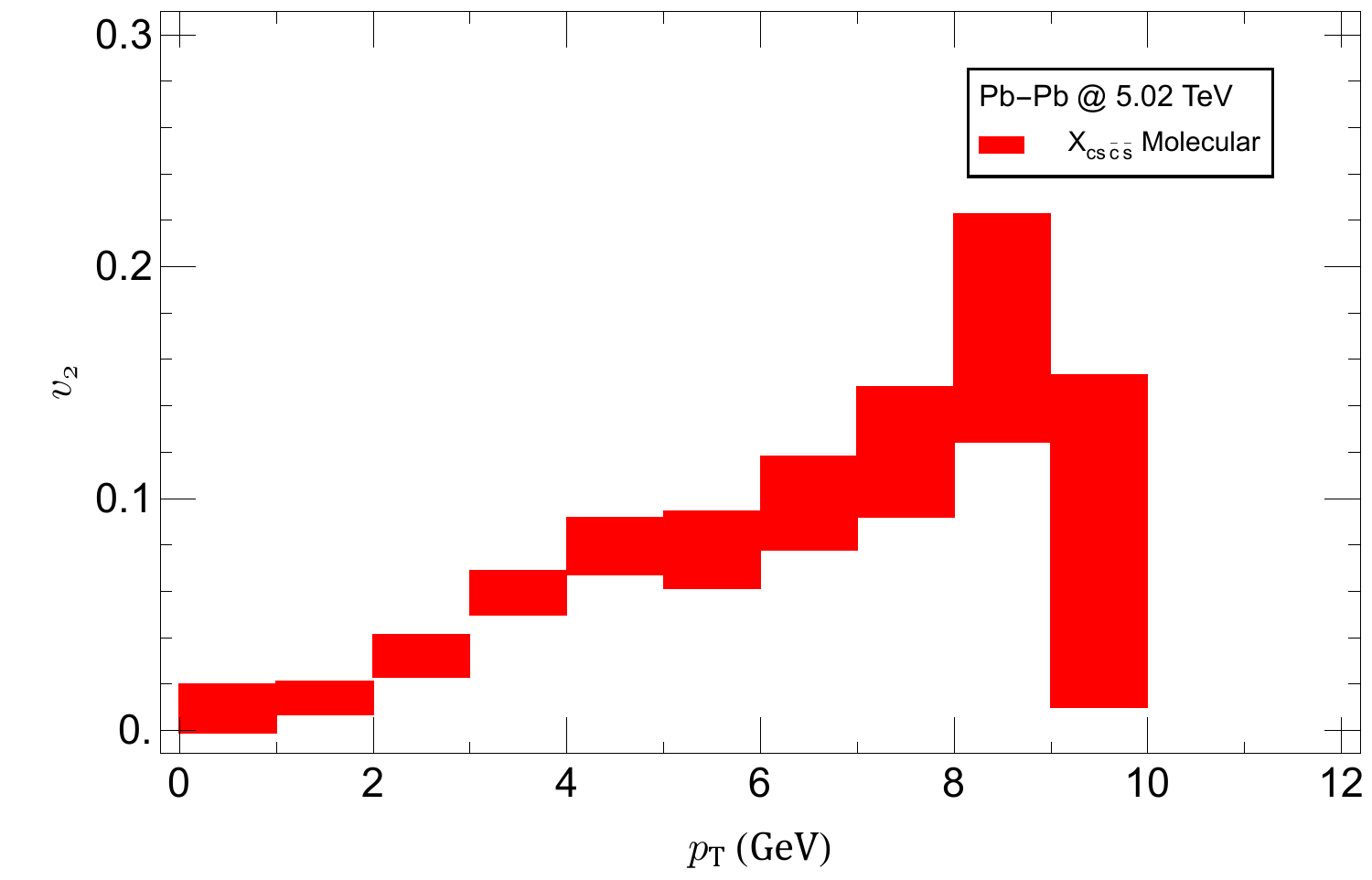}
\caption
{(color online) Elliptic flow coefficient $v_2$ versus transverse momentum $p_T$ for produced $X_{cs\bar{c}\bar{s}}$ in minimum bias Pb-Pb collisions at $\sqrt{s_{NN}}=5.02~\mathrm{TeV}$, predicted from our computation for the hadronic molecule picture. The bands are similarly determined as described in Fig.~\ref{fig:centrality}.} 
\label{fig:elliptic}
\end{figure}

In Fig.~\ref{fig:rapidity}, we present the rapidity and the transverse momentum distributions of $X_{cs\bar{c}\bar{s}}$. One can find that the distribution for both the hadronic molecular state and the tetraquark state is similar to that of the usual hadrons~\cite{CMS:2011aqh, ALICE:2013jfw}. We also show the elliptic flow coefficient $v_2$ of $X_{cs\bar{c}\bar{s}}$ as a function of the transverse momentum $p_T$ in Fig.~\ref{fig:elliptic}. The elliptic flow is sensitive to the geometry of the initial fireball and the generation mechanism of $X_{cs\bar{c}\bar{s}}$. 

\section{Summary and Outlook}\label{sec:Summary}

In this work, we studied the yields of $X_{cs\bar{c}\bar{s}}$ for Pb-Pb collisions at $\sqrt{s_{NN}}=5.02~\rm{TeV}$ by introducing the production mechanism of two possible configurations, i.e., the hadronic molecular state and tetraquark state, into the AMPT model. We found that the production in the molecular picture exceeds in the tetraquark picture by two  orders of magnitude. The centrality distribution of the yields of $X_{cs\bar{c}\bar{s}}$ shows a strongly decreasing trend for the hadronic molecular state and a mild change for the tetraquark state. This system size dependence could be a good probe for the inner structure of $X_{cs\bar{c}\bar{s}}$. We also showed the rapidity and the transverse momentum distributions of $X_{cs\bar{c}\bar{s}}$ production, as well as its elliptic flow coefficient, as a function of the transverse momentum, which can be tested in the future experimental measurements. In Ref.~\cite{ALICE:2021kfc}, a strangeness enhancement effect in heavy ion collisions was found by ALICE Collaboration, which could be evidence for quark-gluon plasma. We expect a similar effect to be found in the ratio of $X_{cs\bar{c}\bar{s}}$ to $X(3872)$, which will be studied in our future work.

\begin{acknowledgments}
The authors would like to thank Dr. J. Liao, E. Wang, Q. Wang, and H. Xing for the helpful discussion. This work is partly supported by Guangdong Major Project of Basic and Applied Basic Research No.~2020B0301030008, the National Natural Science Foundation of China with Grant No.~12105107, Science and Technology Program of Guangzhou No.~2019050001.
\end{acknowledgments}

%%%%%%%%%%Reference%%%%%%%%%%

\end{document}